\documentclass{iaus}
\usepackage{graphicx}
\usepackage{xspace}

\newcommand{\G}{{\sc Gadget}2\xspace}
\newcommand{\LCDM}{$\Lambda$CDM\xspace}

\def\LaTeX{L\kern-.36em\raise.3ex\hbox{a}\kern-.15em
    T\kern-.1667em\lower.7ex\hbox{E}\kern-.125emX}

\def\hMpc{\ifmmode{\:h^{-1}\,{\rm Mpc}}\else{$h^{-1}\,{\rm Mpc}$}\fi}
\def\hkpc{\ifmmode{h^{-1} \,{\rm kpc}}\else{$h^{-1}\,{\rm kpc}$}\fi}
\def\hMsun{\ifmmode{\:h^{-1}\,M_\odot}\else{$h^{-1}\,M_\odot$}\fi}

\title[Too small to form a galaxy] %% give here short title %%
{Too small to form a galaxy: How the UV background determines the baryon fraction}

\author[Hoeft et al.]   %% give here short author list %%
{M. Hoeft$^1$, G. Yepes$^2$ \and S. Gottl{\"o}ber$^3$}

\affiliation{$^1$Jacobs University Bremen, Germany \break m.hoeft@jacobs-university.de\\[\affilskip]
$^2$Universidad Autonoma de Madrid, Spain \break $^3$Astrophysikalisches Institut Potsdam, Germany}

\pubyear{2007}
\volume{244}  %% insert here IAU Symposium No.
\pagerange{119--126}
\date{?? and in revised form ??}
\jname{Dark Galaxies and Lost Baryons}
\editors{J.I. Davies \& M.J. Disney, eds.}
\begin{document}

\maketitle

\begin{abstract}

  The cosmic ultraviolet background (UVB) heats the intergalactic
  medium (IGM), as a result the gas in dark matter halos below a
  certain mass is too hot to cool within a Hubble time. The UVB
  effectively suppresses the formation of dwarf galaxies. Using high
  resolution cosmological hydrodynamical simulations we show that
  photo heating leads to small baryon fractions in halos below $\sim
  6\times 10^9\:h^{-1}\,M_\odot$, independent of the cosmic
  environment. The simulations are carried out assuming a homogeneous
  UVB with flux densities as given by Haardt\,\&\,Madau (1996). A halo
  may stop to condense gas significantly after the universe is
  reionised, namely when its mass falls below the characteristic mass
  scale set by the photo heating. Assuming a spherical halo model we
  derive this characteristic mass analytically and identify the main
  mechanisms that prevent the gas from cooling in small halos. The
  theoretically derived characteristic mass is smaller than the one
  obtained from observations. Increasing the energy per ionising
  photon by a factor between four and eight would be sufficient to
  reconcile both. This is equivalent to an average temperature of the
  IGM of $\sim 10^4\:{\rm K}$. In this sense the faint end of the
  luminosity function may serve as a calorimeter for the IGM.

\end{abstract}

\firstsection % if your document starts with a section,
\section{Introduction}

The faint end of the luminosity function of galaxies indicates that
either only a small fraction of dark matter halos with masses below a
few times $10^{10}\:h^{-1}\,M_\odot$ (`dwarf halos') contain galaxies,
or that galaxies in dwarf halos show on average a high mass-to-light
ratio (provided that the cold dark matter scenario is a good
description of our universe). It has been argued decades ago that the
IGM is highly ionised by a cosmic UVB. The Ly$\alpha$ forest proves
excellently that only a tiny fraction of hydrogen is neutral. The
energy put into the IGM in each photoionisation event leads to
temperatures about $\sim10^4\:{\rm K}$ at present. We
have shown in a preceding work (Hoeft et\,al., 2006) that assuming the
canonical model of the UVB (Haardt\,\&\,Madau, 1996) the resulting
characteristic mass of dark matter halos with baryon deficiency falls short,
more precisely, it is about $6\times 10^{9}\:h^{-1}\,M_\odot$.
\\

Here, we will show in detail how photo heating hampers the gas
condensation in dwarf halos. To this end we present results from
simulations including cooling and photo heating and set up an
analytical model. We argue that the lower end of the galaxy luminosity
function is essentially a measure for the heat input into the IGM.

\section{Simulations}\label{sec:simu}

\begin{figure}
  \begin{center}
  \includegraphics[width=0.37\textwidth,angle=0]{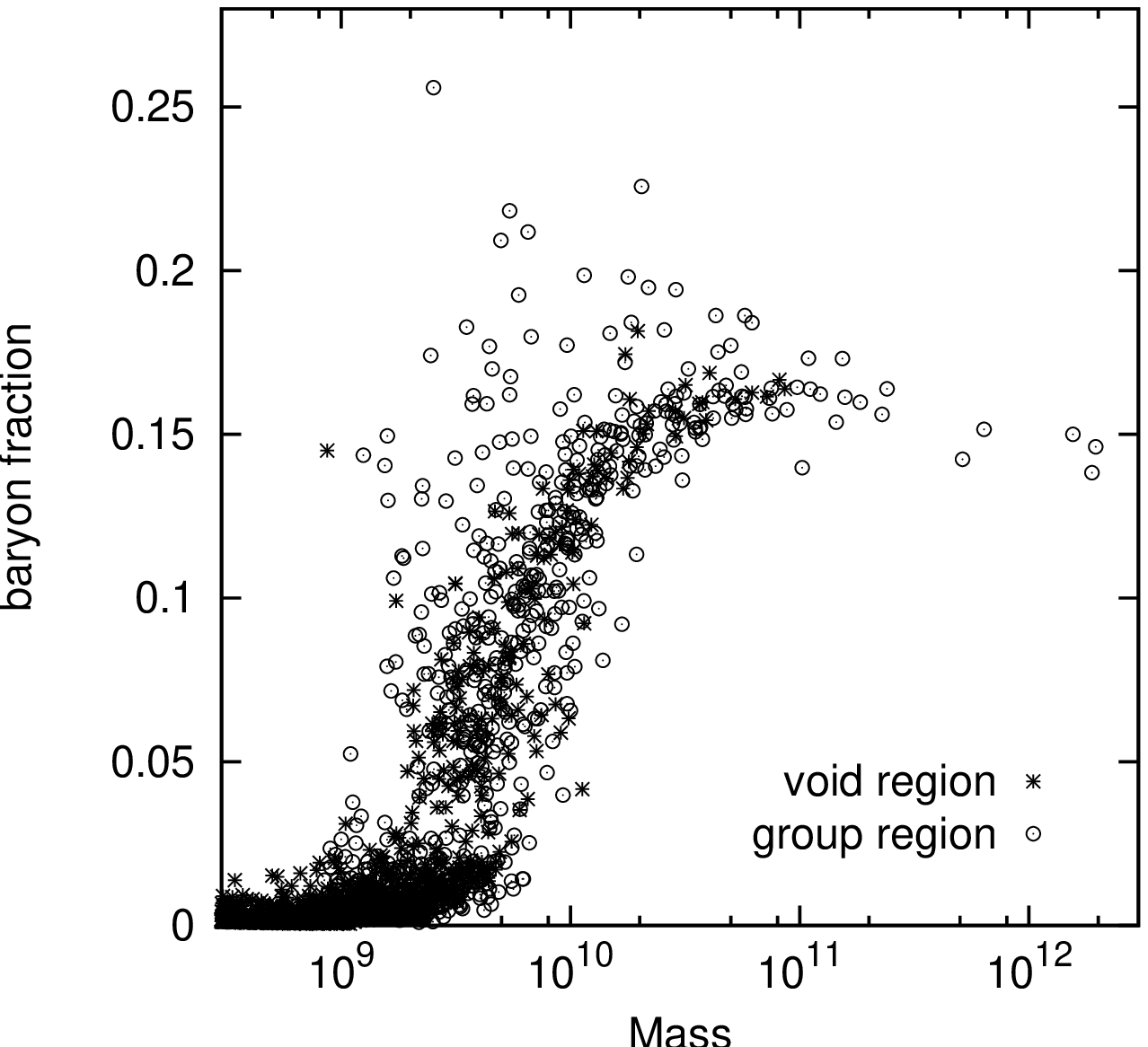}
  \hspace{3mm}
  \includegraphics[width=0.50\textwidth,angle=0]{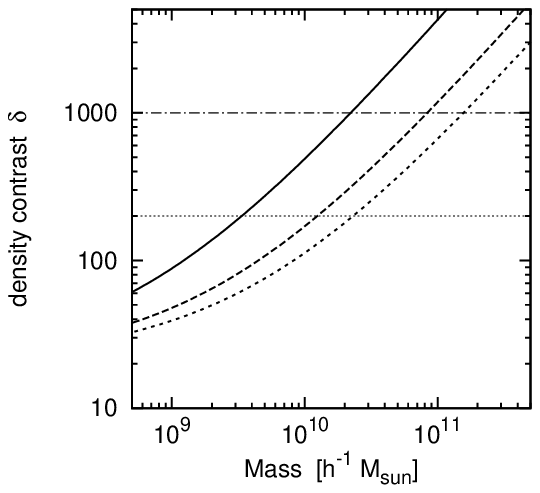}
  \end{center}
  \caption
  {
{\it Left panel:} Baryon fraction within the virial radius. In a cosmological simulation we studied to distinct regions in detail: a void and a loose group region. Only isolated halos are considered, i.\,e. substructures are neglected. {\it Right panel:} The central gas density contrast as a function of total halo mass for three equation-of-states: $T_0 = 3.8\times10^3\:{\rm K}$ (solid line), $8.8\times10^3\:{\rm K}$ (dashed lines), and $1.3\times10^4\:{\rm K}$ (short-dashed line). Moreover, the density contrast at which cooling becomes important is indicated: for $T_0 = 3.8\times10^3\:{\rm K}$ and $8.8\times10^3\:{\rm K}$ at $\delta \sim 200$, and for $T_0 = 1.3\times10^3\:{\rm K}$ at $\delta \sim 1000$. The crossing between the central density contrast and these horizontal lines give the characteristic mass.
  }
\end{figure}

Our simulations have been run with the parallel $N$-body TreePM code \G (Springel, 2005). The hydrodynamical equations are solved using a smoothed-particle-hydrodynamics method based on a entropy-conserving scheme. Radiative cooling is included, assuming a primordial mix of hydrogen and helium. Rates for collisional ionisation, recombination, and cooling are used as given in Katz et\,al. (1996). The gas is heated by a homogeneous UVB radiation. We have adopted a slightly modified version of the photoionisation and heating rates given in Haardt\,\&\,Madau (1996). Star formation and stellar feedback are treated in the code by means of a sub-resolution model in which the gas of the interstellar medium (ISM) is described as a multiphase medium of hot and cold gas. Note, stellar feedback is included in our simulation, but it can only modify the temperature and density of the ISM, we deliberately do not consider galactic winds to focus on the effects of photoheating solely. 
\\

Using a multi-mass technique, we simulate a cosmological void region
and the environment of a loose, poor galaxy group selected from a
computational box of side-lengths $L= 50\:\hMpc$. To construct
suitable initial conditions, we first created a random realization
with $2048^3$ particles using the \LCDM power spectrum of
perturbations of the concordance model ($\Omega_m = 0.3$,
$\Omega_\Lambda = 0.7$, $\Omega_b = 0.04$, $h=H_0 / (100 \, {\rm km \,
  s^{-1} \, Mpc^{-1}} ) = 0.7$ and $\sigma_8 = 0.9$). However, later
on this very high resolution is only used in the regions of interest.
The large-scale gravitational field is represented by more massive
particles. The void was simulated with the full resolution, i.\,e.
$2048^3$, while the group was simulated with $1024^3$.
\\

We compute for each isolated halo in the two regions the baryon
fraction within the virial radius, $f_{\rm b} = (M_{\rm star} + M_{\rm
  gas})/(M_{\rm star}+M_{\rm gas}+M_{\rm dm})$, see Fig.~1, left
panel. One can clearly see that the baryon fraction decreases for
masses below $\sim 6\times 10^9\:h^{-1}\,M_\odot$ (at this mass
$f_{\rm b}$ is half of the average cosmic baryon fraction). Hoeft
et\,al. (2006) showed that this mass scale is very robust even when
changing the resolution, or the UV background flux density over a
large range of parameters. Fig.~1 also indicates that the
characteristic mass is the same for the void and the group region.
Moreover, our simulations allow us to study the evolution of the
baryon fraction in individual halos. Fig.~2 shows four examples. For
more massive halos the evolution of the dark and the baryonic matter
runs virtually parallel. In contrast, less massive halos stop to
increase their content of condensed baryons, i.\,e. stars and cold,
dense gas, at some redshift. That time coincides very well with the
time when the halo mass falls below the characteristic mass, $M_{\rm
  c}(z)$.
\\

The evolution of the individual halos shows clearly how photoheating
acts on the baryon fraction: The gas is heated in a way, that cooling
times in the centre of a halos get longer, hence condensation cannot
take place anymore. The time when this happens is not necessarily
close to reionisation, instead, depending on the mass and the mass
accretion history of a halo, it may also happen close to $z=0$. In the
next section we describe a spherical model which allows to derive the
characteristic mass analytically.

\begin{figure}
  \begin{center}
  \includegraphics[width=0.8\textwidth,angle=0]{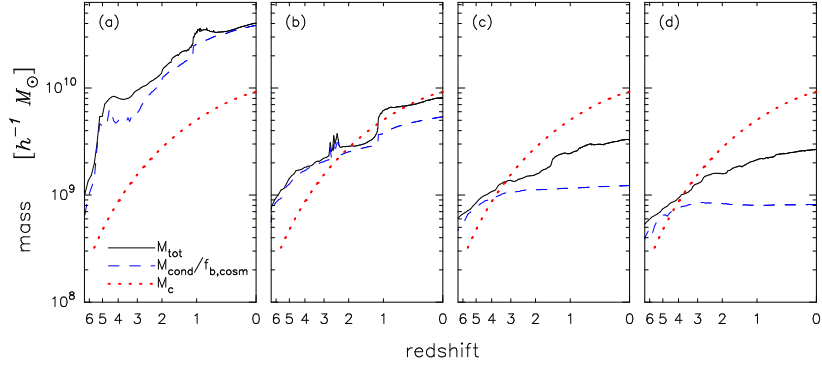}
  \end{center}
  \caption
  {
The mass accretion histories for four halos. In each panel the total mass (solid line) and the condensed mass (dashed line) of one halo is shown. Note, the latter is divided by the mean cosmic baryon fraction, $f_{\rm b,cosm}$, hence, the naive expectation is that both curves lay on top of each other. This allows to recognise easily where condensation starts to fail, namely when the total mass falls below the characteristic mass, $M_{\rm c}(z)$, (dotted line). 
  }
\end{figure}

\section{Why do halos fail to condense gas?}

We wish to derive the mass scale below which halos fail
to condense gas in a simple model. As a first step we could adopt the
effective equation-of-state found for the low density IGM from
numerical simulations. However, we derive this equation explicitly by
integration the thermal evolution of the IGM including atomic hydrogen
and helium species, \{H{\sc\,i}, H{\sc\,ii}, He{\sc\,i}, He{\sc\,ii},
He{\sc\,iii}\} and adopting a primordial helium mass fraction, $Y_{\rm
  p}=0.24$. All species are assumed to be always in collisional
equilibrium. This allows us to follow standard procedures for
computing the thermal evolution of the IGM.
\\

Cosmic structures form out of an almost homogeneous matter
distribution an so do galaxies. If we wish to follow the thermal
history of a small part of IGM we have to know how the density, or
alternatively the density contrast $\delta = \rho_{b}/\bar\rho_{b}-1$,
evolves from the mean density in the beginning to a certain final
density. More precisely, we wish to find a function which describes
the average density contrast evolution, $\delta(t,\delta_0)$, as a
function of the final density contrast, $\delta_0$. In Hoeft et\,al.,
(in prep) we present a fitting formula for the evolution of the gas
density contrast found in the simulations.
\\

The resulting temperature distribution, $T(\delta+1)$, reproduces well
the effective equation-of-state found in numerical simulations. Fig.~3
shows the resulting temperature distribution for several UVB models.
All of them show a power-law for the low density regime, $ T = T_0
(\delta + 1 )^{\gamma-1}$.  For the standard Haardt\,\&\,Madau (1996)
model we find $T_0 = 3.8\times 10^3 \:{\rm K}$ and $\gamma = 1.6 $.
$T_0$ increases slightly if we increase only the heat input at high
redshift. Our maximal model corresponds to a ten times increased 
energy per ionising photon for all redshifts. For this rather extreme
scenario the temperature, $T_0$, rises to $1.5\times 10^4 \:{\rm K}$.
Beside the effective equation-of-state at low density our integration
yields also the transition to thermal balance at higher densities. At 
even higher densities the temperature lays exactly at the border
between heating and cooling, see Fig.~3.  We can read out of
Fig.~3, right panel at which density contrasts the transition to the
thermal balance occurs. For the standard heating model this is at
$\delta \sim 200$ (maximum of the dashed line). In contrast for UVB
models with eight to ten times more energy per ionising photon this
shifts to $\delta \gtrsim 1000$ (dash-dotted and solid line, respectively).
\\

Let us return to the question why massive halos can accrete gas and
small ones cannot. Since photo heating sets the lowest temperature
possible for a given density one can rephrase the questions as
follows: Which halos can compress the central gas sufficiently to
bring the gas into the regime of thermal balance, i.\,e. to make
cooling times short? To answer this question we compute the gas
density profile for a spherically symmetric halo.  It is important to
notice that with an power-law equation-of-state the gas reaches a
finite pressure in the centre of a halo. As we are interested in
rather baryon-poor halos a NFW density profile is a reasonable
approach for the dark matter halo, which dominates the gravitational
force.  Fig.~1, right panel shows the central gas density contrast as
a function of halo mass. Assuming the standard UVB model for halos of
a few times $10^9\:h^{-1}\,M_\odot$ a density is reached where gas
enters the realm of thermal balance, i.\,e. cooling times are short.
The mass scale found by integration agrees with that found in the
simulation. For more energy per ionising photon a higher halo mass is
necessary to reach thermal balance.

\section{More UV heat}

The characteristic mass scale for suppressing gas accretion in dark
matter halos is significantly lower than the mass scale of galaxies at
which the mass-to-light ratio increases dramatically (v.d.\,Bosch
et\,al., 2007). Therefore, one can ask the question, which heat input
is needed to rise the characteristic mass to a level consistent with
observations.  We have increased the energy per ionising
photon in our heating model by a factor of four and by a factor of
eight. In the first case one lowers in general the central density of
the halos, see Fig.~1. In the second case one needs additionally a
higher central density to reach short cooling times, cf. also Fig.~3.,
right panel.  Therefore, the second case represents virtually an upper
limit for the heat input into the IGM. This model would lead to a
temperature $T_0 \sim 1.3 \times 10^{4}\:{\rm K}$ and a characteristic
mass above $10^9\:h^{-1}\,M_\odot$.

\section{Conclusions}\label{sec:concl}

We have presented high-resolution simulations including radiative
cooling and photo heating. Our simulations allow us to determine
robustly the characteristic mass scale below which photo heating
reduces the baryon fraction in dwarf galaxy size dark matter halos.
Using a Haardt\,\&\,Madau (1996) model for the UV background the mass
scale amounts to $M_{\rm c} = 6\times10^9\:h^{-1}\,M_\odot$. The
simulations show that halos start to fail accreting gas when the mass
of the halo becomes smaller than the characteristic mass $M_{\rm
  c}(z)$.

The failure of condensing gas can be traced back to the cooling time
in the halo centre, which may exceed the Hubble time. We have set up a
spherical model for the gas density profile. Crucial for this
model is the effective equation-of-state which we have derived by
integrating the thermal history of the IGM using an approximation for
the density contrast evolution. This spherical model allows us to
derive the characteristic mass for different heating histories of the
IGM.
\\

Therefore, the lower mass scale at which galaxy formation fades away
may serve as calorimeter for the temperature of the IGM. We find that
a heat input with a resulting temperature of $T_0 \sim 10^4\:{\rm K}$
would be consistent with a characteristic mass scale of few times
$10^{10}\:h^{-1}\,M_\odot$. Note that the temperature quoted above may
differ to some extent from that obtained from the Ly$\alpha$ forest
since the heat input in the vicinity of even small galaxies might be
higher than in those structures probed by the forest.
\\

Finally, the combination of cooling times and the effective
equation-of-state shows a remarkable feature, see Fig.~3, right panel:
If $T_0 \gtrsim 1.5 \times 10^4\:{\rm K}$ the equation-of-state runs
quasi parallel to the cooling contour lines. Hence, if a galaxy can
heat all surrounding gas to this temperatures it will stop to accrete
gas for a long time, i.e., radiative feedback may strongly affect the
baryon content in a halo.

\begin{figure}
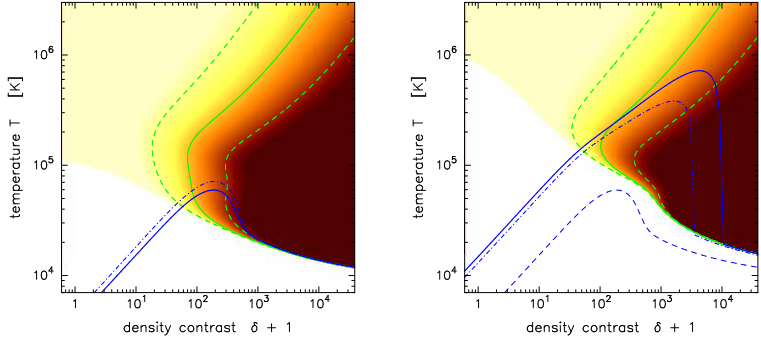

  \begin{center}
  \includegraphics[width=0.33\textwidth,angle=-90]{UV_cooling-standard}
  \hspace{5mm}
  \includegraphics[width=0.33\textwidth,angle=-90]{UV_cooling-fact10}
  \end{center}
  \caption
  {
The temperature, $T(\delta+1)$, distribution for several heating models. {\it Left panel}: Standard UVB heating model (solid line) and four times increased energy per ionising photon only for $z>2$ (dash-dotted line). One can see the effective equation-of-state at low densities and the thermal balance at high densities. The colour in the background indicates the cooling times. Contours are drawn at 10x (dashed line), 1x (solid line), and 0.1x (dashed line) the Hubble time (from the left to the right). {\it Right panel}: Beside the standard model (dashed line) also eight (dash-dotted line) and ten (solid line) increased energy per ionising photon is shown. The cooling times are her computed for the last model.
  }
  \label{fig-Tn-uv-models}
\end{figure}

\begin{acknowledgments}
Numerical simulations has been conducted at the Barcelona Supercomputer Center (BSC) and the CLAMV at Jacobs University, Bremen. This work has been supported by the Deutsche Forschungsgemeinschaft (DFG) under Vo 855/2.
\end{acknowledgments}


\begin{thebibliography}{}


\bibitem[van den Bosch et\,al., 2007]{bosch:07}
     {{van den Bosch}, F.~C., {Yang}, X., {Mo}, et\,al.}
     \textit{MNRAS} 376, 841

\bibitem[Haardt \& Madau, 1996]{haardt:96}
     {{Haardt}, F. \& {Madau}, P.} 1996,
     \textit{ApJ} 461, 20
     
\bibitem[Hoeft et\,al., 2006]{hoeft:06}
     {{Hoeft}, M., {Yepes}, G., {Gottl{\"o}ber}, S. \& {Springel}, V.} 2006,
     \textit{MNRAS} 371, 401
     
\bibitem[Katz et\,al., 1996]{katz:96}
     {{Katz}, N., {Weinberg}, D.~H. \& {Hernquist}, L.} 1996,
     \textit{ApJS} 105, 19

\bibitem[Springel, 2005]{springel:05}
     {{Springel}, V.} 2005,
     \textit{MNRAS} 364, 1105


\end{thebibliography}
\end{document}